# High-frequency Optimally Windowed Chirp rheometry for rapidly evolving viscoelastic materials: application to a crosslinking thermoset

Thanasis Athanasiou[1,2], Michela Geri[3], Patrice Roose[4], Gareth H. McKinley[3] and George Petekidis[1,2]

[1] Institute of Electronic Structure & Laser, FORTH, Heraklion, 70013 Greece

[2] Department of Materials Science and Technology, University of Crete, Heraklion, 70013 Greece

[3] Department of Mechanical Engineering, Massachusetts Institute of Technology, Cambridge MA 02139, United States

[4] allnex Belgium SA/NV, Anderlechtstraat 33, Drogenbos, 1620, Belgium

## Abstract

Knowledge of the evolution of mechanical properties of the curing matrix is of great importance in composite parts or structure fabrication. Conventional rheometry, based on small amplitude oscillatory shear is limited by long interrogation times. In rapidly evolving materials, time sweeps can provide a meaningful measurement albeit at a single frequency. To overcome this constraint we utilize a combined frequency and amplitude-modulated chirped strain waveform in conjunction with a home-made sliding plate piezo-operated (PZR) and a dual-head commercial rotational rheometer (Anton Paar MCR 702) to probe the linear viscoelasticity of these time-evolving materials. The direct controllability of the PZR resulting from the absence of any kind of firmware and the microsecond actuator-sensor response renders this device ideal for exploring the advantages of this technique. The high frequency capability allows us to extend the upper limits of the accessible linear viscoelastic spectrum and most importantly, to shorten the length of the interrogating strain signal (OWCh-PZR) to sub-second scales, while retaining a high time-bandwidth product. This short duration ensures that the mutation number ($N_{Mu}$) is kept sufficiently low, even in fast curing resins. The method is validated via calibration tests in both instruments and the corresponding limitations are discussed. As a proof of concept the technique is applied to a curing vinylester resin. The linear viscoelastic (LVE) spectrum is assessed every 20 seconds to monitor the rapid evolution of the time- and frequency-dependence of the complex modulus. Comparison of the chirp implementation, based on parameters such as duration of experiment, sampling frequency and frequency range, in a commercial rotational rheometer with the PZR provides further information on the applicability of this technique and its limitations. Finally, FTIR spectroscopy is utilized to gain insights on the evolution of the chemical network while the gap-dependence of the evolving material properties in these heterogeneous systems is also investigated.

**INTRODUCTION**

Accurately capturing the linear viscoelasticity (LVE) of a rapidly time-evolving material is an analogue of high-speed photography of a moving object, where exposure times need to be kept short to avoid "motion blur". Likewise, rheometry requires that the measurement of the response of a material is completed in a duration $t_{exp}$ that is short compared to the time-scale characterizing the LVE evolution (Mours and Winter 1994) which is rationalized by the mutation time, i.e.

$$\tau_{Mu}^{|G^*|}(\omega, t) = |G^*(\omega, t)| / \frac{\partial |G^*(\omega,t)|}{\partial t} \qquad (1)$$

where the magnitude of the complex modulus |G*| is assumed to be the parameter of interest. Notably, the value of $\tau_{Mu}$ may be frequency ($\omega$) and age ($t$) dependent. Despite significant progress in instrumentation, current commercial rotational rheometers still interrogate material LVE with "conventional" Dynamic Frequency Sweeps (DFS) within the small amplitude oscillatory shear (SAOS) framework. The sample is excited by a series of stress or strain sine waves of fixed amplitude. The response is acquired sequentially in time as the frequency of oscillation is varied in discrete steps, yielding corresponding data points, which form a discrete LVE spectrum. There are two timescales related to this experiment. On the one hand the period of oscillation $t_{osc} = 2\pi/\omega$ sets the observation timescale for each data point. When the value of $t_{osc}$ is long compared to $\tau_{Mu}$ the assumption of translational time invariance collapses invalidating the measurement. In this case the requirement that strain history response is additive, the so called Boltzmann superposition principle (Markovitz 1977), is not guaranteed (Fielding *et al.* 2000). Hence the evolution of the material under test should be slow compared to $t_{osc}$ for each data point to be valid. On the other hand, the total duration of the frequency sweep sets a longer timescale $t_{sweep}$, (see Figure 1(a)) which governs the coherence of the LVE spectrum. When measuring, for instance, a metastable system each data point at a different frequency, corresponds to a different material age hence, to a material with a potentially different characteristic relaxation time (Kawasaki and Tanaka 2014, Negi and Osuji 2010). Depending on the number of discrete frequencies probed, the duration of the whole frequency sweep experiment can last from a few minutes up to a few hours. To this end, a stricter requirement is imposed for the whole LVE spectrum to be meaningful: the evolution time scales of the material

under test at all frequencies and times tested, expressed by $\tau_{Mu}(\omega,t)$, should be large compared to the duration of the experiment, $t_{exp}$ (in this case equal to $t_{sweep}$). The rate of time evolution for a given experiment can be quantified by the mutation number $N_{Mu} = t_{exp}/\tau_{Mu}$ where values $N_{Mu} \leq 0.15$ have been found to be sufficient to successfully follow the time evolution of the material (Curtis *et al.* 2015).

The stress response of a range of soft materials, including many industrial products and model systems, exhibit quite rapid time evolution owing for example, to changes induced by flow history (Larson and Wei 2019), physico-chemical evolution of the number of active bonds (Macosko 1985, Snijkers *et al.* 2017), crystallization during cooling (Reiter and Strobl 2007), micro-phase separation (Bates and Fredrickson 1990), evaporation (Lehéricey *et al.* 2021), shear-induced phase transition (Parisi *et al.* 2020) and other microstructural changes. This is particularly true for chemically reacting systems such as thermosets. Here, the LVE properties of the crosslinking precursor change nearly exponentially with time when the percolation limit is reached, as a result of the densification of junction points in a three-dimensional space-spanning cluster (Flory 1953). Vinylester resins follow a free-radical polymerization process after addition of a thermal radical initiator in combination with an accelerator in order to allow curing at ambient conditions. Nowadays, vinylester resins (Biron, 2013) are gaining popularity in industrial applications and marine construction and repair. They stand between epoxy and polyester-based resins as they combine the superior mechanical strength, adhesiveness, substantial resistance to chemicals and minimal water permeability of epoxies with the ease of use of polyester-based systems. The final product can be a vinylester fiber-reinforced composite or a vinylester/fumed silica putty with tunable fracture toughness (Adachi *et al.* 2008). The superior resistant of such composites to water permeation makes them the material of choice for boat hulls that are almost "immune" to water degradation (Du Plessis 2010), known as "osmosis" in the relevant technical community.

In general, the evolution of shear modulus of the curing matrix is very important as it provides: a) insights into the network density and the curing kinetics; b) information on the material's processability; c) indications on the presence of residual stresses as a result of volume changes during curing (Crasto and Kim 1993, O'Brien *et al.* 2001) and d) the material property data required to calculate fracture energy and stress propagation length in fiber reinforced networks (Lin *et al.* 2014) of various degrees of cure. The rate of LVE evolution can be quite fast in these systems depending on conditions such as temperature and chemical formulation. In order to address this, many techniques involving alternatives to pure sinusoidal (single wave) excitation have been developed. In the multiwave excitation method (Holly *et al.* 1988) a linear superposition of sinusoids is utilized; therefore the sample is interrogated simultaneously at a number of individual test frequencies. While this decreases $t_{exp}$ substantially, it may easily violate the linearity assumption as the total excitation (strain) amplitude increases considerably, due to the constructive superposition of the different sinusoidal waveforms that are combined. On the other hand, a frequency-modulated excitation waveform, known as a chirp, corresponds to a continuously-swept range of frequencies with a specified lower and an upper limit, $\omega_{min}$ and $\omega_{max}$, respectively, while the amplitude remains constant. Chirps, named after the sound that birds make, are abundant in nature (Wijers 2018) given that any pulse may "chirp" as it propagates in a dispersive medium i.e. due to frequency-dependent dispersion the phase component of the pulse will gradually transform to a chirped waveform. Light pulses propagating in a transparent medium will also chirp due to the accumulated effects of chromatic dispersion and other nonlinearities. Chirps are widely used nowadays in many applications such as sonars and radars (Klauder *et al.* 1960). Fourier-transform based rheometry (Wilhelm 2002) evolved naturally towards exponentially-spaced chirp functions, and the approach has been validated and applied to many soft materials (Curtis *et al.* 2015, Ghiringhelli *et al.* 2012) and has been advanced further by the optimal amplitude modulation (Geri *et al.* 2018). The use of exponential chirps has also been applied orthogonally in conjunction with steady shearing deformation (Rathinaraj *et al.* 2022), to interrogate time evolving material properties and structural anisotropy under flow.

In modern rheometers the advanced motor-transducer assemblies augmented with smart electronics, have significantly improved the data quality and extended the operability range of conventional DFS (Läuger *et al.* 2002). User-commanded oscillatory and transient strain profiles are consistently imposed as long as the input settings remain within the design parametric envelope. The applicability of a combined amplitude-frequency modulated exponential chirp was demonstrated in a conventional rotational rheometer, ARES G2, by using the arbitrary wave option, (Geri *et al.* 2018). Nevertheless, the utilization of these complex strain waveforms in commercial rheometers possesses technical challenges. The ability of these firmware-controlled motors to impose chirps requires instrument specific protocols. This can be tricky given the complexity of modern electromechanical instrumentation and their sample-biased response when operated at the limits of their operational envelope. Our homemade high-frequency piezo-rheometer (PZR) offers a more flexible alternative. Its simplicity and fast response (Athanasiou *et al.* 2019) renders this instrument an ideal test-bed for any arbitrary excitation function and therefore also for applying exponential chirps. Its upper high frequency limit of $\omega = 6000$ rad/s extends significantly the accessed frequency range and, most importantly, allows the use of very short-time interrogating chirps of the order of 1 s.

The aim of this work from the technical perspective is twofold: firstly to explore the merits and limitations of imposing chirp in a sophisticated (i.e. controlled by software with high degree of complexity) dual-head commercial rotational rheometer, here an Anton Paar MCR702; and secondly, to validate high frequency chirp protocols on the PZR. As a platform demonstration we then utilize the optimally windowed chirps (OWCh) to study the rapidly evolving LVE spectrum of our material test system, a commercial vinylester resin under cure. This OWCh protocol has been applied successfully to measurements in a commercial rheometer before (Geri *et al.* 2018, Keshavarz *et al.* 2021) as well as in simulations (Bantawa *et al.* 2023, Bouzid *et al.* 2018) however, to the best of our knowledge this is the first implementation in an Anton Paar MCR 702 and also in a sliding-plate piezo-based high frequency configuration such as our PZR. We also aim to determine the optimized operational window as well as the limitations of the technique in both instruments based on their design characteristics.

**LINEAR FOURIER TRANSFORM RHEOMETRY: FROM DFS TO OPTIMALLY WINDOWED CHIRP (OWCH)**

In a conventional strain-controlled DFS test, the frequency dependence of the complex modulus ($G^*(\omega)$) is derived by correlating the measured stress response waveform $\sigma(t)$ of the material under test to the harmonic strain excitation $\gamma(t) = \gamma_0 \sin(\omega_i t)$ for each angular frequency measured at a given strain amplitude $\gamma_0$. Here we use the index $i = 1, 2,\ldots, N_f$ to denote each frequency step where $N_f$ is the total number of frequency points tested. The determination of the discrete LVE spectrum from the time-domain signals is expressed mathematically,

$$G^*(\omega_i) = \frac{\hat{\sigma}(\omega_i)}{\hat{\gamma}(\omega_i)} \tag{2}$$

where the caret denotes the Fourier Transform of the relevant time time-dependent signal. In this single-wave (harmonic) excitation (Figure 1(a)) and within the linear limit, the material will respond with a sinusoidal stress waveform $\sigma(t) = \sigma_0 \sin(\omega_i t + \delta)$ i.e. the material responds at the same frequency (as the imposed deformation) with a phase advance $\delta$ related to its viscoelasticity. A continuous signal obeying the Dirichlet conditions related to convergence (Lanczos and Boyd 2016), can be decomposed into a superposition integral of exponentials of infinite duration by the Continuous Fourier Transform (CFT) (Bachman *et al.* 2000). However, in real measurements the excitation and response signals are discrete functions (i.e. sequences of numeric values, indexed on integer variables). The sampling time which is the reciprocal of the sampling frequency $f_s$ and signal duration $t_{exp}$ are both finite, hence the Discrete Fourier Transform (DFT) is the relevant transformation to the frequency domain. The DFT maps a number of $N = f_s t_{exp}$, equally spaced in time, real values (data) to $N/2 + 1$ equally spaced discrete frequencies (bins) in the Fourier space. In this framework the DFT will represent a discrete (digitized) time signal, shear stress for instance, with a Fourier series by calculating its projection onto an orthogonal basis set such as complex exponentials or the more intuitive sine ($\delta = 0°$) and cosine ($\delta = 90°$) waveforms at discrete frequencies $\omega_i$

$$\sigma[t_n] = \sum_{i=0}^{N/2} a_i \sin[\omega_i t_n] + b_i \cos[\omega_i t_n] \qquad (3)$$

where the sample index $n = 0, 1, 2..N$, increments the time in discrete steps $t_n = n/f_s$, and the frequency of the analyzing Fourier series $\omega_i = i \cdot \frac{2\pi f_s}{N} = i \Delta\omega$ i.e. integer multiples of the frequency spacing $\Delta\omega$. Equation 3 describes the reconstruction of a time series from its Fourier bases and is shown in its most intuitive form with the brackets indicating the discrete nature of the sampled spectrum. Notably the first component $\omega_0$ is zero, a DC component, not directly related to LVE of the material and always rejected by the rheometric software. In our data this is also ignored. Nonetheless its value could reflect residual stresses from the prior strain history. When the two Fourier coefficients $a_i$ and $b_i$ are normalized by the applied strain amplitude in the spirit of Equation 2 they yield the in-phase component of stress scaled with strain, i.e. the storage modulus $G'(\omega_i)$ and the quadrature term (i.e. the 90° shifted component) representing the loss modulus $G''(\omega_i)$. In most applications the DFT is computed via Fast Fourier Transform (FFT) which is actually a class of algorithms capable of reducing computational effort from being proportional to $N^2$ to $N \log_2 N$. The FFT algorithm cleverly assumes a number of calculations as redundant exploiting symmetry (Oppenheim *et al.* 2001). The most prominent limitation of the DFT stems from the relation $\Delta \omega = \frac{2\pi f_s}{N} = 2\pi/t_{exp}$ i.e. the measurement duration trades-off frequency resolution. Consequently an optimum combination of $f_s$ and $N$ is needed for a given application. The spectral contribution of a given signal located between two bins inevitably is smeared over both of them inducing errors, referred to as spectral leakage. Another characteristic of the DFT inherited from the CFT of continuous signals is the assumption that the waveform is infinite in duration. Likewise, the DFT treats any discrete input signal as periodic with infinite duration (Harris 1978).

Multiwave deformation signals can also reduce $t_{exp}$ as the excitation acquires simultaneously information at multiple frequency data points (Holly *et al.* 1988). Equation 2 holds for the superposition of any number of sine waves as well as for any arbitrary excitation signal, provided the total strain amplitude remains within the LVE limit. This constraint restricts the number of excitation sine waves (frequencies) applied, as the different components will eventually superimpose constructively. This is clearly indicated in Figure 1(b) where four waves with $\omega_1$ = 1 rad/s, $\omega_2$ = 4 rad/s, $\omega_3$ = 14 rad/s and $\omega_4$ = 53 rad/s are superimposed. The maximum strain amplitude of this strain waveform with a period of $2\pi/\omega_1$ = 6.28 s is 4% compared to 1% associated with each component. An ideal multiwave signal would contain all frequencies of interest in a single waveform whilst keeping the amplitude envelope constant. A frequency modulated signal known as the *chirp* fulfills these requirements. It has been utilized in rheometry as an excitation signal (Ghiringhelli *et al.* 2012) with the method referred to as Optimal Fourier rheometry. The chirp can be conceived as a sine wave of the form $\gamma(t) = \gamma_o \sin[\varphi(t)]$ with an instantaneous frequency that grows in time (up-chirp), as

$$\omega(t) = \frac{d\varphi(t)}{dt} = \omega_{min} \left(\frac{\omega_{max}}{\omega_{min}}\right)^{\frac{t}{t_{exp}}} \qquad (4)$$

where the instantaneous frequency $\omega(t)$ is deduced from the time-derivative of the phase angle $\varphi(t)$ and $t_{exp}$ is the total duration of the chirp signal.

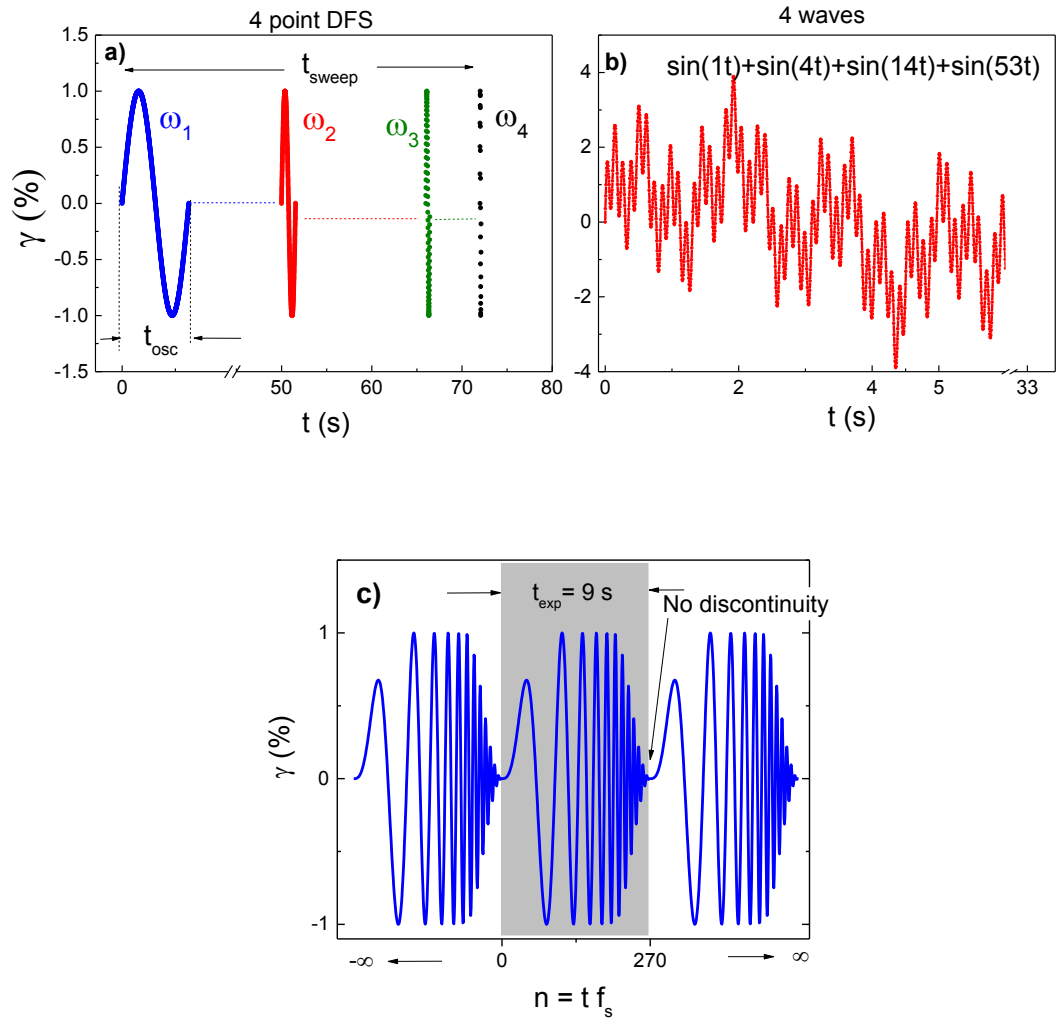

Figure 1. Strain waveforms utilized in three popular rheometric techniques for measuring LVE properties: a) DFS at four frequencies: $\omega_1 = 1$ rad/s, $\omega_2 = 4$ rad/s, $\omega_3 = 14$ rad/s and $\omega_4 = 53$ rad/s. Only 1 period of each signal is shown for clarity. The total duration of the experiment is $t_{sweep}$ while $t_{osc}$ sets the time scale for the acquisition of each frequency point. b) Multiwave excitation with a superposition of the same frequencies each with $\gamma_0 = 1\%$. The resulting wave has a period of $2\pi/\omega_1 = 6.28$ s but the maximum strain amplitude is 4%. One period is shown while the x axis is broken to indicate the duration of 5 periods of the total signal. c) An optimally windowed chirp pulse (OWCh) ranging from 1 to 30 rad/s (grey shaded area). The signal is rendered periodic by infinite repetition of the sequence to past and future times. The tapering of the bounding signal envelope is set to 50% to illustrate the absence of discontinuities as the chirp amplitude at the beginning and the end of the pulse reduce smoothly to zero. Chirp duration is 9 s. Abscissa is the sample index with $N = 270$ samples and $f_s = 30$ Hz.

In conventional DFS, the test frequency is varied by applying a sequence of sine-waveforms, each typically few periods long, with frequencies ranging from $\omega_1$ to $\omega_{N_f}$. This is illustrated in Figure 1(a) for a case of $N_f = 4$ (with $\omega_1 = 1$ rad/s, $\omega_2 = 4$ rad/s, $\omega_3 = 14$ rad/s and $\omega_4 = 53$ rad/s ) where only one period of each signal for the four point DFS is shown for clarity. The typical duration (acquisition time) of this test when using the MCR702 rheometer in the default

Page 8

setting is 70 s approximately, as multiple periods are utilized by the instrument. The reduction of the number of acquired oscillations at each frequency can shorten the total test time. The DFT input is a sinewave (for both stress and strain of Eq. 1) in discrete form, reducing the transform to a multiplication of the discrete signal with a sine and cosine reference time series of the same frequency. Here there is only one non zero (known) expected output frequency (bin), thus with proper frequency spacing and accurate sampling (timing) spectral leakage becomes non-existent. As the DFT acts like a combination of band-pass filters (Prabhu 2014), in this case a single one, all noise and harmonics are rejected and only the base frequency is considered. On the other hand, in a chirp signal the DFT is computed once for the whole spectrum hence, only frequencies smaller than $\omega_{min}$ and larger than $\omega_{max}$ can be ruled out. The chirp in the grey shaded area of Figure 1(c) is $t_{exp} = 9$ s and $N = 270$ samples long. The sample number N can be increased by either faster sampling rates or increased chirp duration. The inherent periodicity of the DFT extends this sequence, with infinite repetitions (cloning), into the past and into the future. The amplitude tapering i.e. the gradual increase and decrease of the amplitude of the chirp beginning and end respectively (see Figure 1(c)) removes discontinuities that induce harmonics and eventually spoil the whole spectrum by spectral leakage.

By reducing the observation time $t_{exp}$ we reduce the information acquired and most importantly we increase the effect of spectral leakage known also as Fresnel ripples (Kowatsch and Stocker 1982, Oppenheim, Buck and Schafer 2001). Spectral leakage is thus the smearing of a frequency component or relaxation mode in rheological terms, to adjacent frequencies. The vulnerability of the resultant LVE spectrum to spectral leakage can be quantified by the dimensionless time bandwidth product (Levanon and Mozeson 2004):

$$TB = t_{exp}(\omega_{max} - \omega_{min})/2\pi \qquad (5)$$

Equation (5) is easily conceived from the fact that both the signal acquisition time $t_{exp}$ and the frequency range affect the frequency resolution. Leakage is mitigated considerably by increasing TB and by tapering the amplitude at the beginning and end of the chirp by a suitable windowing (weighting) function such as the widely used Hann, Hamming, Tukey or other window (Harris 1978, Nuttall 1981). The choice of the windowing function depends on the application (Prabhu 2014). An optimally windowed chirp is a jointly frequency and amplitude modulated waveform with proper tapering to avoid unnecessary loss of power at the edges of the spectrum. Tapering is quantified by the tapering ratio $r$ which ranges from 0 to 1 as defined in the Appendix. A tapering ratio $of$ $0.06 < r < 0.15$ was found to be best by (Geri *et al.* 2018) hence the name Optimally Windowed Chirp (OWCh). Based on this study we will utilize a tapering ratio $r = 0.1$ (10% of the signal duration) and the Tukey window function (Tukey 1967) throughout our experiments.

**EXPERIMENTAL SETUP**

*Rheometric geometries*

The two basic geometries utilized in the piezrheometer (PZR) and the commercial rotational rheometer (MCR 702) are first briefly reviewed. Whatever the excitation signal may be, the transfer function of the sample $G^*(\omega)$, is interrogated by means of a strain actuator and a stress sensor, referred to in the language of commercially rotational rheometers as the motor and transducer, respectively.

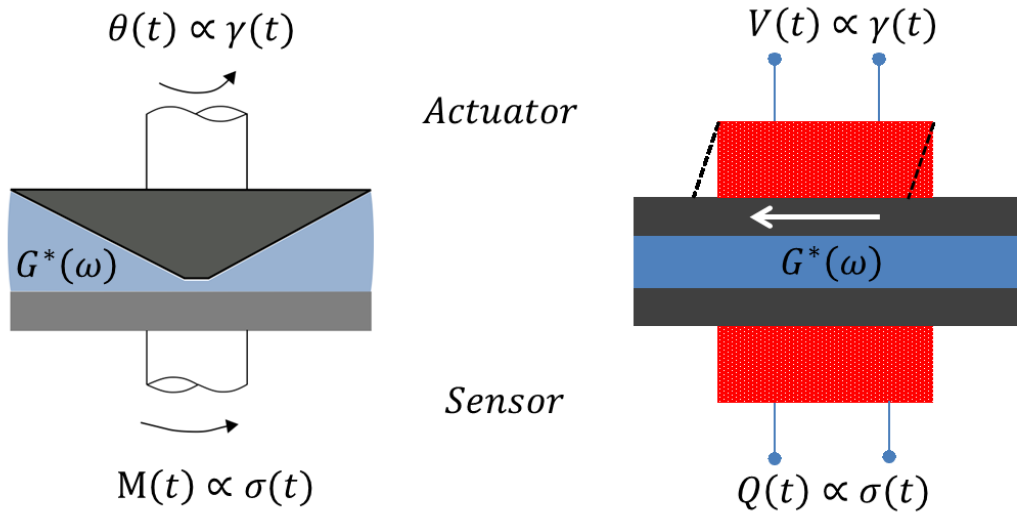

Figure 2 Geometries utilized in the MCR 702 and PZR i.e. cone-plate and sliding-plate rheometer respectively. A separate motor transducer (SMT) arrangement and strain controlled mode are depicted. In each case the upper plate is the moving surface and the lower is the sensing surface: a) Cone and plate fixture in a rotational rheometer; angular deflection $\theta(t) \propto \gamma(t)$ is the controlled parameter (excitation) while torque response $M(t) \propto \sigma(t)$ is the measured quantity (response). b) Parallel-plate or sliding- plate geometry based on piezoelectric ceramics depicted in red. The voltage $V(t) \propto \gamma(t)$ is the controlled parameter while the total charge generated $Q(t) \propto \sigma(t)$ is the measuring quantity. The dimensions are not-to-scale.

A typical strain control setup is visualized in Figure 2 for two of the most popular geometries:

*a) The cone and disk geometry (cone-plate) found in commercial rotational rheometers:* For small cone angles the strain excitation, $\gamma(t)$, can be approximated as being proportional to the angular deflection, $\theta(t)$, of the upper tool connected to the motor. The stress response is determined by the lower tool which is connected to the transducer and is proportional to the torque signal, $M(t)$, (Macosko 1994). In the Anton Paar MCR 702, motor and transducer consist of identical components.

*b) The elementary sliding plate geometry utilized in our homemade high frequency piezo-rheometer (PZR):* Here rectilinear deformation is imposed by design. A voltage waveform $V(t)$ excites the piezo actuator to induce a proportional strain, $\gamma(t)$, while the resulting stress is followed by the lower piezo which acts as a shear stress sensor via the proportional generated charge, $Q(t)$. The plates used are circular with a radius of 7.5 mm and the gap is set to 200μm to minimize sample inertia effects.

Both setups comprise separate motor transducer (SMT) arrangements in which tool inertia effects are minimal (Athanasiou *et al.* 2019). While the operational principles are the same, there are differences mainly as the user has no control on the sophisticated MCR 702 firmware. The firmware/ software control the motor rotation and the stress acquisition loop. Sliding plate geometries are not often encountered in commercial rheometers as they cannot impose steady shear. They suffer less though from edge effects and sample trim errors, as the sample near the interface does not contribute to the total integrated torque as much as the annular strip of sample located at the rim of a rotational geometry (Cardinaels *et al.* 2019).

*Data acquisition and parametric envelope*

In the present work all the experiments are strain-controlled. The OWCh discrete strain signal is initially generated in tabulated form, by means of modifications to the original MIT OWCh Matlab code (see Equation A1 at the Appendix) (Geri *et al.* 2018) and then delivered to the instrument. Both the MCR702 and PZR are commanded to apply a series of *N* strain steps $[t_n]$, with $t_n = n\Delta t$ in sufficiently small time steps $\Delta t = 1/f_s$. The stress response, $\sigma[t_n]$, is monitored with the same sampling frequency $f_s$ and the LVE spectra is calculated from equation (2) by means of the MIT OWCh code. For the strain signal to be approximated as continuous and avoid parasitic harmonics we require $f_s \gg \frac{\omega_{max}}{\pi} =$ Nyquist sampling rate (Nyquist 1928) and this conditions is often referred to as oversampling. This is vital to prevent aliasing (when higher frequencies components are observed at a lower i.e. *aliased* frequency) and to reduce high order harmonics due to quantization. The technical limitations of each instrument impose restriction on the maximum $f_s$ and $N$. In the Anton Paar MCR702, the data points for both strain and stress are limited to $N_{max}$ = 2000 per interval, as more data points cannot be handled efficiently by the instrument's digital circuitry resulting is loss of data. The maximum acquisition rate is $f_{s,max}$ = 200 Hz. By maximizing $f_s$ and $N$, the digitization error is reduced and the corresponding chirp duration $t_{exp} = N/f_s$ = 10 s. This optimum combination of $N$, $f_s$ and $t_{exp}$ is summarized in Table 1. To eliminate the possibility of data loss we safely performed most measurements in the MCR702 with $t_{exp}$ = 9 s resulting in $N$ = 1800 samples at the maximum acquisition frequency.

Table 1: Limitations related to data acquisition for MCR702 and PZR. The accuracy of the measurement is optimum for each instrument when the maximum values are utilized. Oversampling is the ratio of sampling frequency over $4\pi\omega_{max}$ where $\omega_{max}$ is the maximum frequency tested denoted in brackets. Signal amplitude resolution is not known in MCR 702.

|        | $N_{max}$ (samples) | $f_{s,max}$ (kHz) | Oversampling      | Amplitude resolution (max) |
|--------|---------------------|-------------------|-------------------|----------------------------|
| MCR702 | 2k                  | 0.2               | 6 (100 rad/s)     | -                          |
| PZR    | >200k               | 800               | 420 (6000 rad/s)  | 12bit ($2^{12}$ steps)     |

The PZR strain actuator and stress sensors are piezoelectric ceramics allowing measurements at frequencies up to 6000 rad/s. We modified the original setup (Athanasiou *et al.* 2019) by adding a multichannel digital to analogue (D/A) and analogue to digital (A/D) converter, MicroDAQ 2000 (Embedded solutions, Poland). Contrary to the MCR702 rheometer, here the circuitry is fast enough to process data in real time while $N$ is restricted only by the computer's memory capacity. The stress response is digitized after conditioning of the signal by means of a charge amplifier (essentially a current integrator) (Horowitz *et al.* 1989, Starecki 2014). The output of the charge amplifier, $V(t)$, is proportional to the generated charge $Q(t)$ and the rheometric parameter of interest i.e. the stress, $\sigma(t)$. Since the dependence is linear, the conversion from voltage (V) to stress units (Pa) is accomplished via calibration. Both D/A and A/D converters are set to a minimum of 12 bit resolution corresponding to a quantization in $2^{12}$ steps, to enhance amplitude resolution and minimize induced harmonics. We refrained from exploiting the maximum resolution of 16 bit of the A/D unit in order to minimize the size of the acquired data. The sampling frequency can be as high as 800 kHz an order of magnitude higher than in the MCR702. To perfectly synchronize $\sigma[t_n]$ and $\gamma[t_n]$ the latter (imposed) signal was re-sampled by an additional A/D channel. Failure to perform such synchronization would result in a few milliseconds delay of the actual excitation compared to the commanded one and this can induce a considerable phase angle error at the higher frequencies of oscillation. This error was initially detected by our calibration method described below. Following this correction, the FFT of the strain and stress signals were calculated by means of the OWCh MIT code and then the real and imaginary components of the complex modulus (G´ and G´´) were determined from Equation 2.

The OWCh signal can be as short as the period of the lower frequency component contained i.e. $2\pi/\omega_{min}$. On the other hand, as the signal duration $t_{exp}$ is increased, $TB$ increases and the deviation from the FFT assumption of infinite duration is mitigated resulting in decreased spectrum leakage. Although the limit is not rigorous, a value of $TB \geq 100$ is preferred (Kowatsch and Stocker 1982). Figure 3 summarizes the operability window given all the operating parameters and limiting factors discussed. $TB$ is calculated from Equation 5 for the PZR and MCR (red and olive dashed line respectively) based on: a) the requirement to keep mutation number low such as $N_{Mu} = 0.2$; this maps the acquisition time $t_{exp}$ (upper x axis) to the mutation time (lower x axis); b) the usable frequency range i.e. 1 to 60 rad/s and 10 to 1000 rad/s for the MCR702 and PZR respectively. For the MCR 702 frequency range can be extended to lower values ($\omega_{min} < 1$ rad/s) with subtle effect on TB while frequencies higher than 60 rad/s induce significant errors in the imposed strain waveform. Also due to restrictions in $f_s$ and $N$ (see Table 1) the signal is optimized at a duration of $t_{exp} = 10$ s indicated with the olive arrow. Larger $t_{exp}$ will improve the time-bandwidth product $TB$ at the expense of the sampling frequency $f_s$. The PZR can be operated at $t_{exp} = 0.7$ s, nevertheless the record length $t_{exp}$ has sufficient margin to be increased for the benefit of $TB$, when $\tau_{Mu}$ is large. The grey shaded area marks the area where $TB$ falls below 100 for both instruments. In this regime the obtained LVE is prone to spectral leakage and results should be assessed with caution. We stress here that this limit is not strict as proper windowing and increased sampling rate can mitigate spectral leakage.

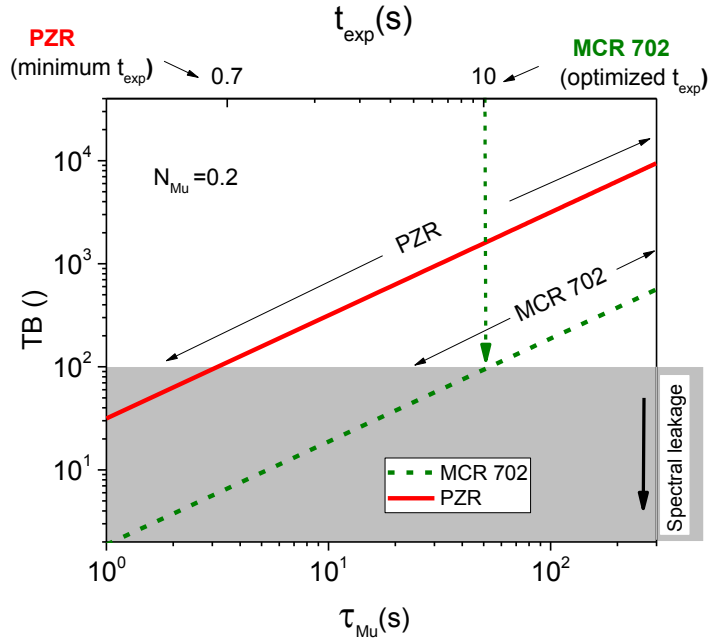

Figure 3. Operability window for exponential chirps in the MCR 702 and PZR under the requirement that $t_{exp} = \tau_{Mu}/5$ and therefore $N_{Mu} = 0.2$. Hence $t_{exp}$ (upper x axis) is mapped to $\tau_{Mu}$ (lower x axis). Dashed green and red solid lines depict the calculated $TB$ values for MCR702 and PZR from Equation 5 with $\omega_{min} = 1$ and 10 rad/s and, $\omega_{max} = 60$ and 1000 rad/s respectively (see also text). Grey shaded area indicates regime with increased spectral leakage. The green dashed arrow indicates the optimized $t_{exp}$ for MCR702.

**VALIDATION IN MCR702**

Modern rotational rheometers excel in conventional tests such as DFS. MCR702 performs well even at frequencies as high as 300 rad/s (Athanasiou *et al.* 2019) as its firmware can compensate for motor and transducer inertia in the separate motor transducer mode. However, when commanded to impose a non-conventional excitation such as the OWCh profile the actual realized strain deviates from the commanded (nominal) value mainly in the last few cycles of the chirp which correspond to the highest instantaneous frequencies. This deviation was significantly increased when the instrument was commanded to perform a 5s OWCh with $\omega_{max} = 100$ rad/s (see Fig. S1). This deviation can be viewed as an additional (effective) tapering of the end of the chirp pulse. In this "dummy" mode it seems that the built-in EC motor is not augmented much by the smart electronics and algorithms designed to optimize classic oscillatory deformations (Läuger *et al.* 2002). The maximum sampling frequency $f_{s,max}$ of 200 Hz is six times above Nyquist limit (Table 1). This six-fold oversampling is sufficient to prevent aliasing yet remains a limiting factor to minimize spurious frequency components when the signal is digitally sampled in the presence of noise. The number of points acquired per measurement is also limited to $N_{max}$ = 2000. Given these factors a chirp duration of $9 \leq t_{exp} \leq 10$ s is found to be optimal as it exploits the maximum sampling rate of 200 Hz and mitigates mismatch of commanded and actual strain signals. Additionally $\omega_{max}$ was limited at 60 rad/s. Small deviations of strain amplitude can be taken into account by normalizing the stress response in Eq. (1) with the actual strain logged by the instrument. On the other hand, large distortions of the strain waveform (as in the case of Figure S1) will eventually alter significantly the power spectrum of the signal resulting in the growth of unwanted harmonics. Spectral noise will impair the resolution of the measured spectrum as the signal to noise ratio decreases. Contrary to a conventional DFS, these harmonic distortions may spread along the whole VE spectrum introducing additional spurious relaxation modes that are not related to material properties. The entire LVE spectrum may be affected in this case and not just a few frequency points as in DFS. We therefore limited the chirp experiment to frequencies up to $\omega_2$ = 60 rad/s and $N$ = 1800 samples to improve the overall data quality and avoid overloading the MCR702 data buffer. No data averaging or smoothing was performed in order to explore the limitations of the method.

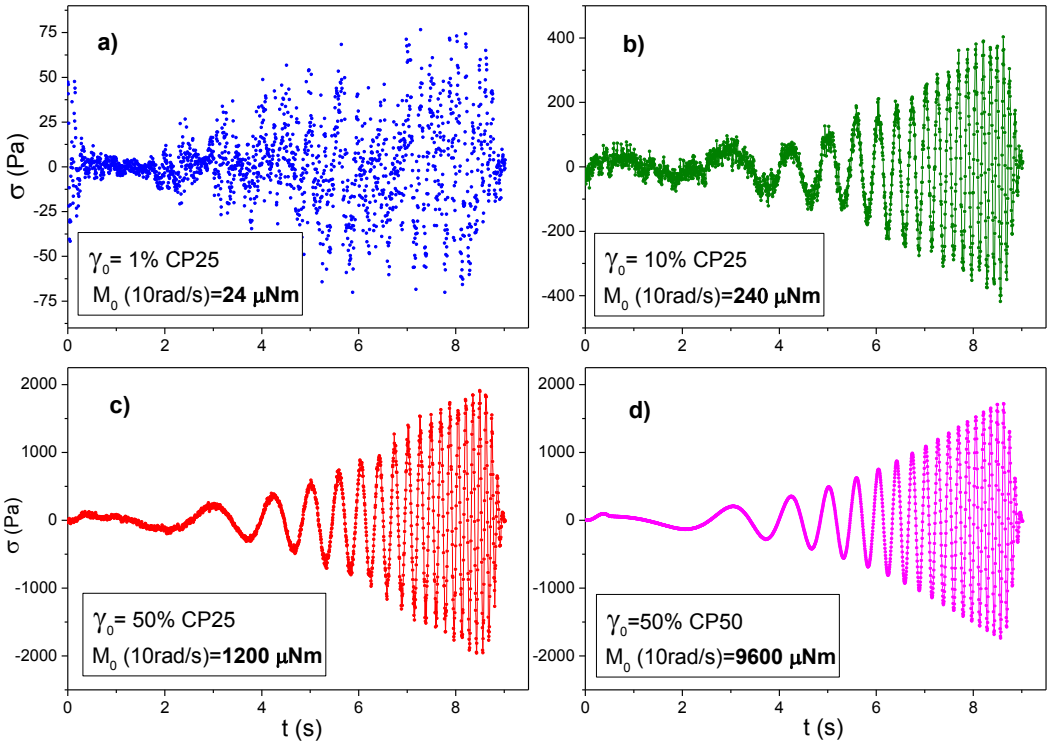

Figure 4. Discrete stress response signals of PDMS 100 Pa s to chirp excitation with $\omega_{min}$= 1rad/s, $\omega_{max}$= 60 rad/s, r = 0.1, $t_w$ = 0.02 s and $t_{exp}$ = 9 s in the MCR702, Anton-Paar rheometer. The legends indicate the strain amplitude, the geometry used and the corresponding torque amplitude ($M_0$) when a 10 rad/s strain oscillation is applied. The corresponding time bandwith product TB = 84.5. The local maximum at early times is due to the weak sample response at lower frequencies and the amplitude tapering parameter $r \neq 0$.

To demonstrate experimentally the effect of the response signal (stress) strength i.e. transducer torque amplitude $M_0$, we performed a series of tests with a PDMS oil of zero shear rate viscosity of 100 Pa s (Brookfield, USA). The excitation chirp parameters $\omega_{min}$ = 1 rad/s, $\omega_{max}$ = 60 rad/s, $r$ = 0.1, $t_w$ = 0.02 s and $t_{exp}$ = 9 s as well as the time-bandwidth product $TB$ = 84.5 (see Eq.5) were kept constant. The strain amplitude $\gamma_0$ and the tool geometry were varied to increase systematically the torque amplitude, i.e. to gradually increase the response signal strength (while ensuring they remain in the linear regime). To rationalize the response signal strength we report in Figure 4 the equivalent torque value, $M_0$, corresponding to a 10 rad/s strain oscillation, at the strain amplitude and geometry specified in the legend. This measure of signal strength can be easily determined by a conventional dynamic frequency or time sweep experiment. At weaker values of the torque signal (with CP25 mm and $\gamma_0$ =1%), the stress response to chirp signal is quite noisy. The signal resolution improves as $\gamma_0$ is increased, with the value of the torque $M_0$ indicated in the legend. As expected the noise in the digitized stress signal affects the whole LVE

Page 16

spectrum (Figure 5(a)) and especially the weaker hence more susceptible to spectral noise elastic modulus $G'(\omega)$. (Figure 5(b)). Satisfactory agreement of OWCh with the conventional DFS derived spectra was achieved only when the strain amplitude was increased to 50% and the experimental scatter in the value of $G'$ was eliminated at even stronger torque signal when the geometry was doubled in radius (CP50mm). A nominal torque amplitude of $M_0$ = 1200 μN m at 10rad/s was thus required for an acceptable measurement. In contrast when the MCR702 executes a conventional DFS with CP25mm a torque amplitude of $M_{10\,rad/s}$ = 24 μN m, two decades larger that the nominal sensitivity limits of the instrument, is sufficient for the measurement. However the conventional DFS takes minutes to complete rather than $t_{exp}$ = 10s. This illustrates the trade-offs inherent to selection of the most appropriate frequency-domain technique for a specified material or problem.

The effect of noise is also evident in the power spectral density (Figure 6) where the whole spectrum is compromised for the smallest strain amplitude $\gamma_0$ = 1%. Tapering reduces power at the higher frequencies in the excitation (black curve) while the measured response signal retains high values at higher frequencies due to the increase of the complex modulus magnitude $|G^*|$ with $\omega$. In conventional DFS this results in the improvement of data quality in moderate to higher frequencies for samples where $|G^*|$ increases with ω. Based on our experiments it appears that the torque amplitude should be at least 10 times higher in a chirp based than in a conventional DFS. The difficulty to provide an unambiguous criterion stems from the fact that the error magnitude is also sample dependent. For instance a weak dependence of $|G^*|$ on frequency or the phase angle, $\delta$, approaching the limits of 0 and 90° where $tan(\delta)$ is sensitive, will increase the error. A prediction of the effects of spectral leakage is also non-trivial. The presence of noise in the measured response signal at any time during the experiment ($0 < t < t_{exp}$) can affect the whole spectrum. This cannot be filtered out with conventional signal processing, as time information is lost in the frequency domain (John *et al.* 2023). The limits and optimal settings discussed in this section refer only to viscoelastic liquid systems measured by MCR 702 and can serve only as a general guidance.

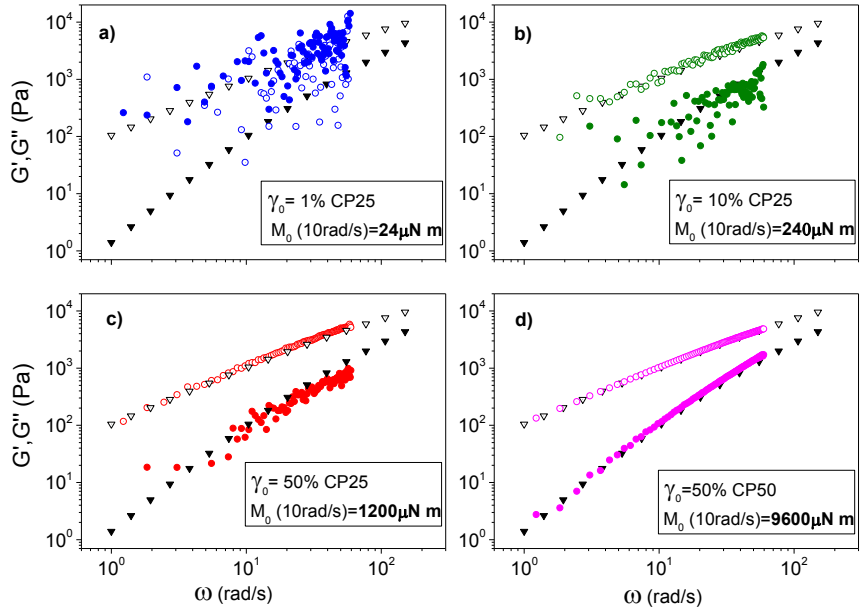

Figure 5. LVE spectra of PDMS 100 Pa s obtained by OWCh excitation at a constant time bandwidth product $TB = 84.5$ in MCR702 at 23 °C. Torque signal strength is gradually increased from a) to d) as indicated in the legend. Black triangles are reference data from conventional DFS from the same instrument. Filled and unfilled symbols denote $G'$ and $G''$ respectively.

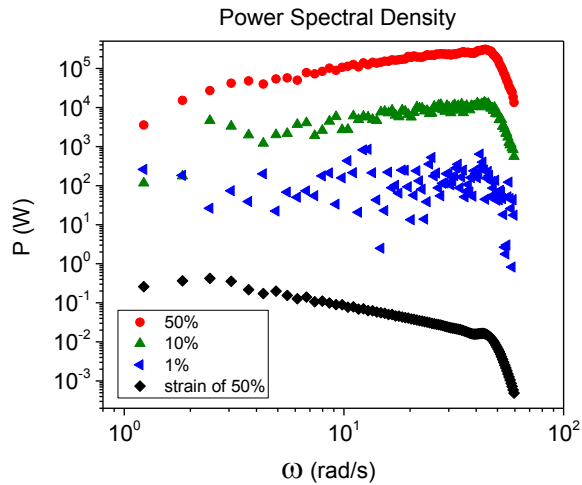

Figure 6: From the top: power spectral density (or spectral energy distribution per unit time (s)) of stress response for the 50, 10 and 1 % strain amplitude. The black curve is the power spectral density of the actual imposed (logged by the instrument) strain at 50% amplitude.

**VALIDATION IN PZR**

The applicability of OWCh-PZR was validated in a frequency range of $10 \leq \omega \leq 1000$ rad/s. High frequency chirp signals can have short durations without compromising the data integrity by keeping $TB > 100$. Even a signal as short as $t_{exp} = 0.7$ s is possible with reduced spectrum leakage effect (Figure 3). To validate the chirp measurements in the PZR, a pDMS and a linear polybutadiene (pBD) solution were measured. The former is a viscosity standard (Brookfield, USA) with nominal zero shear viscosity of 1000 Pa s at 25 °C. LVE reference data were provided by means of conventional DFS-MCR702 up to 300 rad/s. The latter sample is a thermo-rheological simple material hence, hence its LVE spectrum can be time-temperature shifted. This allows conventional rotational rheometers to extend their accessible frequency range to higher frequencies by cooling the sample and shifting the resulting LVE data. Polybutadiene with a molar mass of 44.5 kg/mol, was diluted in squalene at a concentration of 75 w/w%. Cyclohexane was used as a co-solvent with the viscous squalene and subsequently removed under reduced pressure while 2,6 di-tert-butyl-p-cress was added to prevent subsequent oxidation. The sample prepared is a well entangled solution with 19 entanglements per chain (Rubinstein and Colby 2003). Dynamic frequency sweeps were performed at various temperatures from 23 to – 60 °C with an ARES MELT rheometer (with a 2k force transducer FRTN1) at 3% strain amplitude. All data were shifted to a reference temperature $T_{ref} = 23$ °C. The reference LVE data were obtained with this method extending the high frequency limit to $10^5$ rad/s. The same sample was also measured in PZR with the OWCh method. The chirp frequency varied exponentially in time from 10 to 1000 rad/s as shown in Figure 7(a), according to Eq. 3. The whole frequency scan lasted only 0.7 s including an initial 0.2 s of waiting time for the setup and initiation. The excitation and the response, raw, signals are illustrated in Figure 7(b), (c) as voltage waveforms.

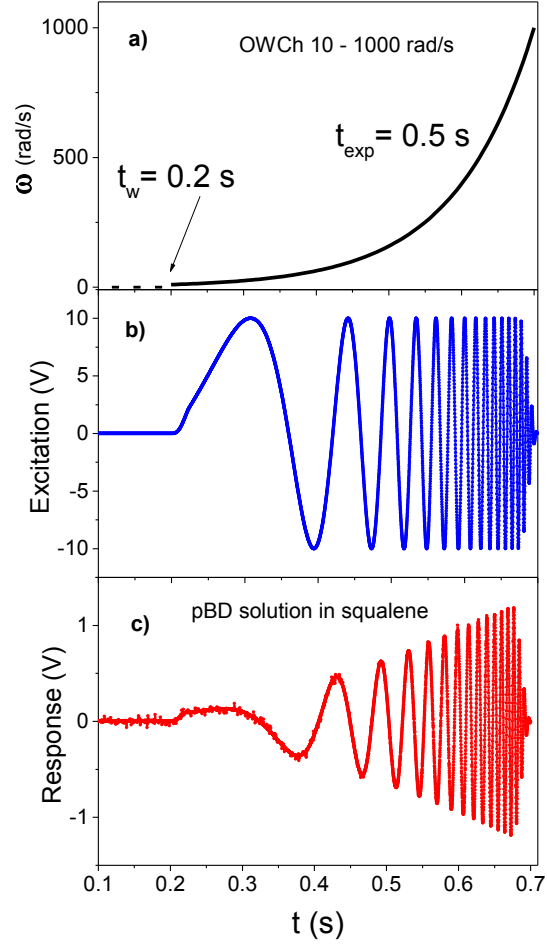

Figure 7. High-frequency excitation chirp (OWCh) and response of polybutadiene solution in PZR. From top to bottom: the exponential variation of instantaneous frequency $\omega(t)$ calculated from Equation 4 varying from 10 to 1000 rad/s, the excitation and the response raw signal in volts. Here $t_{exp}$ = 0.5 s, $t_w$ = 0.2 s, $r$ = 0.1 and $TB$ = 78.

After normalization of the response signal based on the initial PZR calibration and by means of Eq.2, data were converted to the frequency domain. Figure 8(a) depicts the validation in PZR of the OWCh method. Data obtained by the OWCh–PZR method superimpose nicely with that from a conventional frequency sweep in the PZR (denoted DFS-PZR) and reference DFS measurement by MCR702, although the latter data are limited in frequency range. Notably, this is a direct comparison between the DFS-PZR method where data are extracted via a lock-in amplifier for each individual frequency with the OWCh-PZR method in which data are calculated via FFT of a single frequency-modulated time domain signal. The OWCh-PZR method is further validated in Figure 8(b), where the data are now compared with reference data obtained from a conventional rotational rheometer. The frequency range of these reference data is extended by time-temperature superposition as the sample (linear polybutadiene solution) is thermorheological simple. Notably, the OWCh-PZR data were validated at the shorter allowable measurement time, $t_{exp}$ = 0.5 s, corresponding to a marginal value of $TB$=78. Nonetheless, below we will utilize for all our applications a safer value of $t_{exp}$ = 1 s resulting in a twofold increase of $TB$ that minimizes the effect of spectrum leakage.

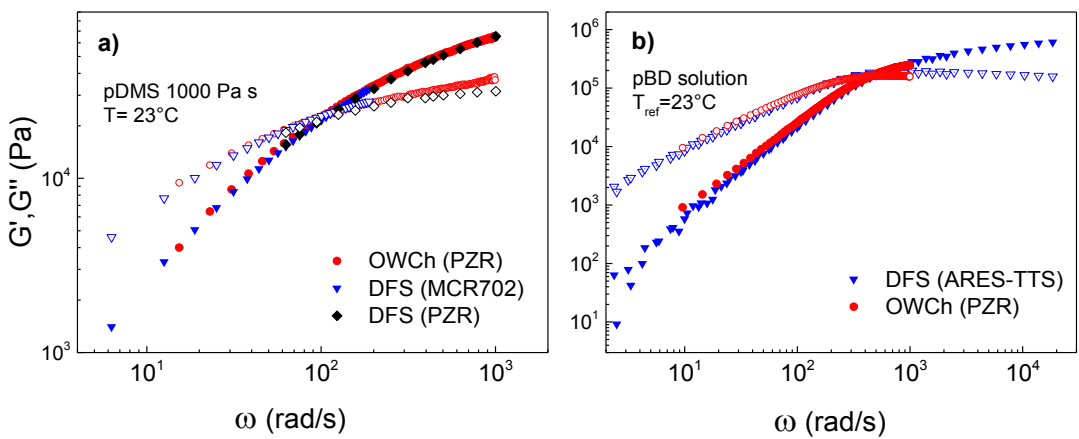

Figure 8. Validation of OWCh-PZR method: a) comparison of LVE spectra of pDMS obtained by DFS-MCR702 (reference), DFS-PZR and OWCh-PZR; b) comparison of LVE spectra of pBD solution in squalene obtained by time temperature shifted DFS-ARES- data (reference), and OWCh-PZR.

**CURING OF A VINYLESTER RESIN (VER)**

VERs refer to a class of unsaturated thermoset resins typically based on methacrylated epoxy resins diluted in styrene. Upon addition of a radical initiator, along with a suitable accelerator (e.g. metal derivatives such as cobalt hexanoate), VER's can "cold" cure i.e. they can crosslink at ambient temperature following a free-radical crosslinking mechanism and hence are widely used in the area of fiber reinforced composites. Peroxide initiators such as methyl ethyl ketone peroxide (MEKP) are commonly used to start the reaction. Once added to the resin however, resin formulations can only be used for a finite time before turning into a gel. As an unsaturated monomer, styrene acts as an efficient reactive solvent which is incorporated in the cured network leaving no by-products when polymerization is completed. For the purpose of this work, crosslinking of a commercial VER sample (HQ800 A45 supplied by Sirca s.p.a., Italy) with a styrene content of 32 w/w% (as measured by gas chromatography-flame ionization detection), was studied. This is a slow curing resin suitable for producing fiber reinforced composites via the vacuum infusion method (Abdurohman *et al.* 2018). Usually a post-curing step (i.e. heating at elevated temperatures after initial curing at room temperature) is required for these resins to achieve their ultimate mechanical strength.

The curing kinetics are both temperature and MEKP concentration dependent (Martin *et al.* 2000); hence, for all of our experiments these were kept constant i.e., T = 23°C and 1.8% w/w respectively, unless otherwise noted. The initiator was systematically introduced in the resin using a roller mixer for 180 s. All times are reported from the time instant of MEKP addition. After additional 180 s of rest, the sample was loaded on to the instrument and the evolution in the time dependent complex modulus $G^*(\omega, t)$ was followed.
Macroscopically, VER's exhibit gelation and vitrification during isothermal curing. In the ideal case the transition between sol and incipient gel (gel point) during polymerization is defined when the phase angle tangent $tan(\delta) = G''/G'$ becomes frequency independent (Chambon and Winter 1987). Here we did not attempt to determine the gel point precisely; however, beyond this point the sample behaves increasingly like a rubbery solid. As the degree of crosslinking further increases, the glass transition temperature ($T_g$) of the curing resin gradually approaches the curing temperature. When $T_g$ crosses the curing temperature, the system starts to vitrify leading to a reduction in free volume and concomitantly severe kinetic constraints for the reacting radicals (Harran and Laudouard 1986, Lange *et al.* 2000). As a result, in this glassy state the curing rate drops significantly and the modulus approaches its final plateau value slowly over time. There are cases where the percolating network deviates from the ideal self-similarity (Aoki *et al.* 2019, Lieleg and Bausch 2007) hence gelation can be attributed to the transition point from liquid to solid-like behavior yet this is frequency dependent. In the following subsection we focus on the post-gel state where the high signal to noise ratio (high modulus results in strong signal) that prevents spurious bands emerging as artifacts in the measured spectrum. Only SAOS data taken with MCR702 are shown for the pre-gel state.

*Viscoelastic response during curing*

We first discuss results with conventional DFS and specifically a short duration DFS consisting of a limited number of discrete data points. This 5-point DFS, of decreasing frequency from $\omega_1 = 200$ rad/s down to $\omega_5 = 1$ rad/s, lasts only 84s. Nevertheless, the elastic modulus data are inconsistent as suggested by the negative slope in Figure 9(a). On the other hand, when $|G^*|$ for each frequency point is plotted against the actual time that was measured, the data become meaningful. Indeed Figure 9(b) shows the time evolution of $|G^*|$ for 4 different frequencies. They all refer to the same loading hence same crosslinking path. The data exhibit two exponential regimes (at early and late times respectively) which is typically observed in many evolving, crosslinking systems (Martin, Laza, Morras, Rodrıguez and Leon 2000). From the different slopes in Figure 9 it is clear that $\tau_{Mu}$ is frequency and time dependent. We will focus here on the faster $|G^*|$ evolution at $\omega = 1$ rad/s that is detected for t > 1500 s. The dashed line indicates the best data fit with a fit function indicated in the inset of Figure 9(b). Following Eq. 1 we obtain a characteristic time $\tau_{Mu} = 101$ s that is constant within this time range, corresponding to a $N_{Mu} = 0.83$, a value larger than the criterion of $N_{Mu} < 0.2$. This clearly suggests that the experimental duration T must be reduced further, below 20 s, values that are however not feasible in conventional DFS even with a highly limited frequency step of a few points per decade.

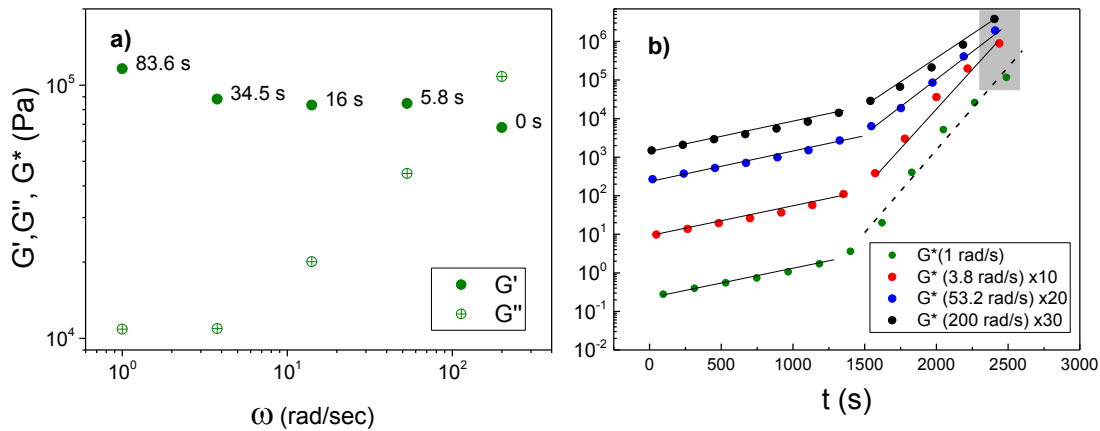

Figure 9. LVE of the crosslinking VER measured with the MCR 702 and a 25mm parallel-plate geometry: a) 5-point conventional DFS give rise to an inconsistency i.e. a negative slope in $G'$ with increasing frequency; numbers indicate the interval time in seconds, i.e. the elapsed time since the 5 point DFS started; b) evolution in the magnitude of the complex modulus $|G^*|$ from a sequence of 12 intervals of 5-point DFS as a function of actual

crosslinking time (since the addition of MEKP) of each frequency point. Only 4 out of 5 different frequencies are shown, shifted vertically for clarity as indicated. Solid lines are guides whereas the dashed line is a best fit to the data shown ($\log|G^*| = 4.29\ 10^{-3}t - 5.39$). Data included in the grey shaded area correspond to the 12th DFS interval depicted in panel (a).

Here the benefit of OWCh rheometry becomes evident. In contrast to a conventional DFS, a $t_{exp}$ = 10 s OWCh-MCR702 measurement is sufficiently short for the proper determination of the LVE of the sample with a mutation number $N_{Mu} \approx 0.1$. Likewise, the 1 s OWCh-PZR measurement corresponds to $N_{Mu} \approx 0.01$ providing nearly isotemporal data at each frequency scan.

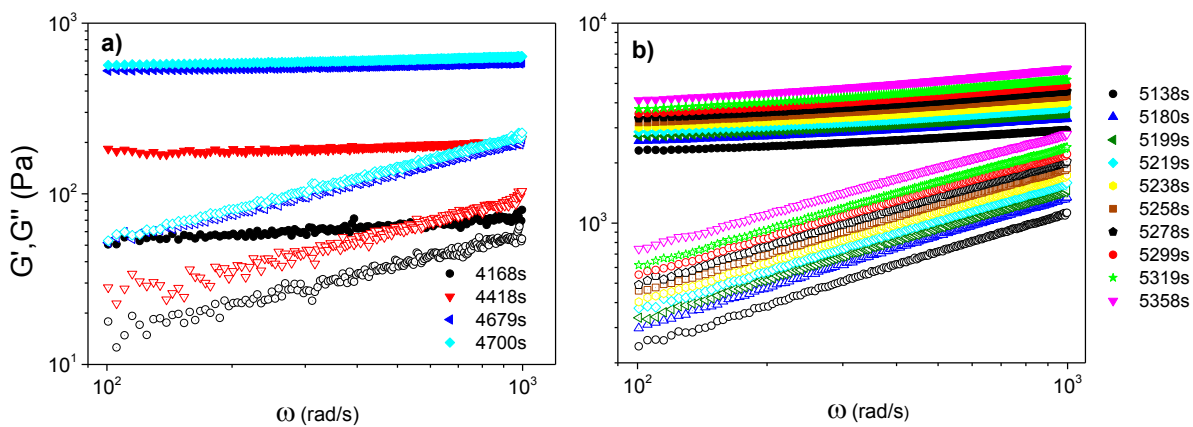

Figure 10. LVE spectra of the crosslinking VER measured with the OWCh-PZR technique. A few selected representative DFS out of the total 56 measurements taken, are shown for clarity at early (a) and long (b) times. Legend indicates the elapsed time from MEKP addition and mixing. $G'$ and $G''$ are represented by filled and open symbols respectively.

To this end, catalyzed vinylester resin (with MEKP 1.8 w/w%) was loaded in the PZR. A series of 56 OWCh measurements were performed with a repetition period of approximately 20 s and $t_{exp} = 1\ s$ at the post gel regime. The spectra obtained from four of these, namely the 1st, 12th, 25th and 26th chirp, at early times, are shown in Figure 10 (a). The 1st OWCh-PZR measurement at 4168 s (70 min) indicates a high frequency crossover of $G'(\omega)$, $G''(\omega)$ slightly above 1000 rad/s, attributed to local intra-chain interactions. In this, sub- millisecond, time scale there is still some freedom for local motion therefore $G''$ approaches $G'$ in analogy with dynamics related to the in-cage rattling of colloidal glasses (Athanasiou et al. 2019) although in the latter, the physics are fundamentally different. As the network densifies this crossover shifts to higher frequencies and further out of the experimental window (red curve at 4418 s), while both moduli increase one order of magnitude within 500 s (blue curve at 4679 s). The cyan curve corresponds to times only 21 s later capturing the significant moduli evolution in this short period. The evolution at long times represented by the last 10 chirps is followed in Figure 10(b) where both moduli continue to increase with $G'$ observed to increase faster at high frequencies consistently with the data of Figure 9. On the other hand, the phase angle exhibits a non-monotonic behavior as seen in Figure 11(a). At early times it constantly decreases as the sample becomes more solid-like whereas at long times it increases again. This increase, indicative of an increased dissipation in the system as curing progresses may appear counter intuitive for a crosslinking network with a constantly increasing junction density.

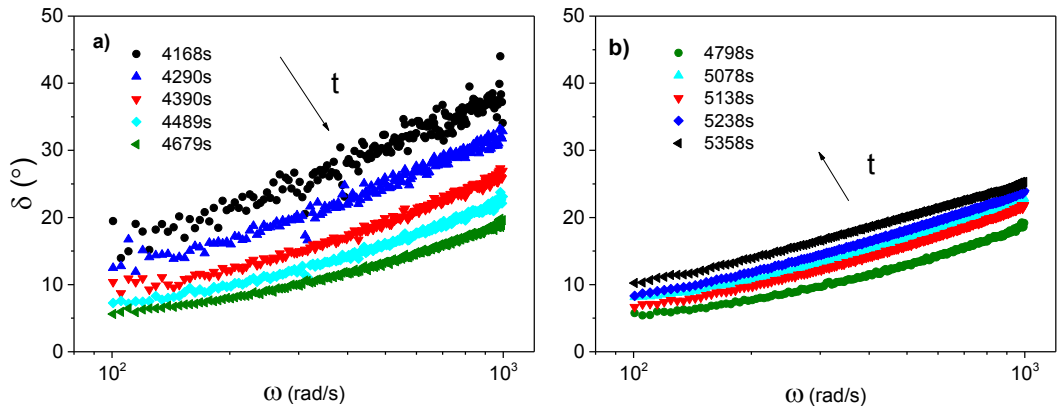

Figure 11. Dependence of the measured phase angle $\delta(t)$ on curing time of vinylester resin measured with the OWCh-PZR tecnique. A few representative frequency sweeps out of the total 56 runs are shown for clarity at early (a) and long (b) times. Arrows indicate the direction of increasing time.

However, LVE properties are also affected by the glass transition temperature, $T_g$ that constantly evolves with curing time hence the distance from $T_g$ differs significantly at each moment (Aoki *et al.* 2019, Harran and Laudouard 1986, Lange *et al.* 1999). Further discussion on this phase angle increase can be found below.

Page 25

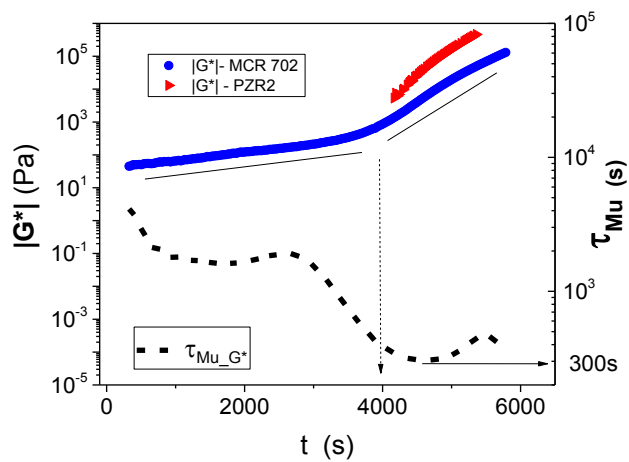

Figure 12. The evolution of $|G^*|$ at 200 rad/s with crosslinking time. Data correspond to DTS with MCR702 (blue spheres) and a sequence of 56 OWCh (red triangles) with PZR where the 200 rad/s point is shown from each spectrum acquired. Lines are guides. Dashed arrow marks the time (4000 s) where the exponential growth of $|G^*(t)|$ changes slope. Dashed curve denotes the mutation time variation as curing evolves. Solid arrow identifies the lowest value of the mutation time scale $\tau_{Mu}^{|G^*|}$. Gap is set to 0.2 and 0.15 mm for the MCR702 and PZR respectively.

The evolution of $|G^*|$ at 200 rad/s, since the onset of crosslinking, measured by dynamic time sweep (DTS) in MCR702 is depicted in Figure 12 where two exponential regimes are probed. Interestingly, the slope in the log-linear plot changes at 4000 s. To determine the origin of these two regimes additional insight into the chemical crosslinking reactions taking place is needed as discussed in the next section. Before going to that we compare $|G^*|$ derived from OWCh-PZR and DTS-MCR702 at the frequency of 200 rad/s that is accessible by both instruments. The moderate discrepancy between the blue and red curve in Figure 12 can be attributed to minor differences in the temperature history during and after mixing. Uncertainty in the exact value of time cannot be entirely eliminated due to different loading procedure between MCR702 and PZR.

Two additional reasons that potentially maybe important are the following: Firstly, the crosslinking path is not identical when an experiment is repeated. This discrepancy is enhanced when the sample dimensions and the dissipation rate of the generated heat are different. Humidity absorption also may play a role (Acha and Carlsson 2013), however this variation is assumed small under laboratory conditions and for the limited duration of our experiments. Secondly, confinement effects become important in the smaller gap of PZR or similarly in small gap plate-plate geometries in conventional rheometers or near the center in a cone-plate geometry (see discussion in Appendix B). Therefore, we used a parallel plate (25 mm) geometry with a gap of 0.2 mm in the MCR 702, which is close to the gap of 0.15 mm of the PZR to enable better comparison between the two data sets. Still however this does not ensure data an exact and quantitative comparison between the two instruments even though the gap is quite similar due to the inhomogeneous microstructure and different history. The dependence of mutation time on curing time can be calculated from Eq. (1) based on MCR702-DTS data as depicted in Figure 12 (blue curve). The minimum of mutation time of 300 s (worst case scenario) dictates $N_{Mu}^{|G^*|}$ < 0.03 and 0.003 during the OWCh-MCR702 ($t_{exp} = 10\ s$) and OWCh-PZR ($t_{exp} = 1\ s$) measurements, respectively. Notably, these values are related to $|G^*|$ as the parameter of interest (hence the superscript) at the frequency of 200 rad/s. Given that this information is not *a priori* known in experiments, working at small $N_{Mu}$ values is important to ensure proper measurement of the LVE of a time evolving material. In industrial applications, this is important considering that most general purpose resins cure considerably faster, usually within 1000 s while fast-curing epoxies require only 300 s to solidify.

*Spectroscopic monitoring of VER curing*

In order to understand the evolution of the network formation that results in the two exponential regimes observed in Figure 12 we probe quantitatively the chemical reactions taking place during curing. The LVE measurements shown in Figure 12 are limited in time mainly due to technical reasons already discussed. However, at the point where the rheological measurements stopped (< 2h) the fairly low values of $|G^*|$ (≈ 0.1 MPa) suggest that the polymerization of the VER is not completed yet. Ultimately, a glassy network should form and, upon vitrification, a ramp in modulus towards ~ 0.1 GPa (a typical value for crosslinked resins) would be expected (Saiev *et al.* 2023) which does not appear in the curing profiles. Information of the viscoelastic transformation can be obtained by monitoring the free-radical polymerization in similar conditions using FTIR spectroscopy. Prior to analysis, the content of methacrylate double bonds of the epoxy methacrylate oligomer of the VER was independently characterized by standard proton NMR. Overall the amount of double bonds prone to radical polymerization in the VER is 4.3 mmol/g with 3.1 mmol/g from styrene and 1.2 mmol/g from the epoxy methacrylate oligomers. However, as a mono-unsaturated monomer, styrene does not introduce crosslinks upon polymerization. Crosslinking follows from the incorporation of the epoxy methacrylate which is telechelic (i.e. a precursor or oligomer with two similar end groups, here methacrylates).

With 1.2 mmol/g of methacrylate double bonds one can anticipate that the VER leads to a moderately dense network and that crosslinks are connected by strands of styrene.

The curing kinetics was followed by time-resolved ATR-FTIR spectroscopy (Chung and Greener 1988) in order to monitor the consumption of the reacting methacrylic and styrenic double bonds. In Figure S2(a), a comparison of the fingerprint region for the FTIR spectra of the VER with and without styrene reveals the strong and specific IR absorption bands of styrene owing to its aromatic nature. Showing little overlap, the band at 908 cm$^{-1}$ was particularly convenient to study the consumption of the styrene double bond as further illustrated in Figure S2(b) where the FTIR spectra of the initial VER and a partially cured state are compared. In contrast, the low amount, sensitivity and spectral differentiation of the methacrylate double bonds offered few options. The small specific absorption peak at 813 cm$^{-1}$ ($\tau(=CH_2)$) was selected for quantification of the double bond consumption along the polymerization. For accurate area determination, a fitting approach of Gaussian profiles was applied when required to solve the overlap with neighboring peaks as highlighted in Fig. S3. Every 60 s, 16 spectra (wavenumber range 650-4000 cm$^{-1}$, resolution 2 cm$^{-1}$) were collected and returned as average in order to improve the signal-to-noise ratio. The total acquisition time for one averaged spectrum was $\approx$ 20 s.

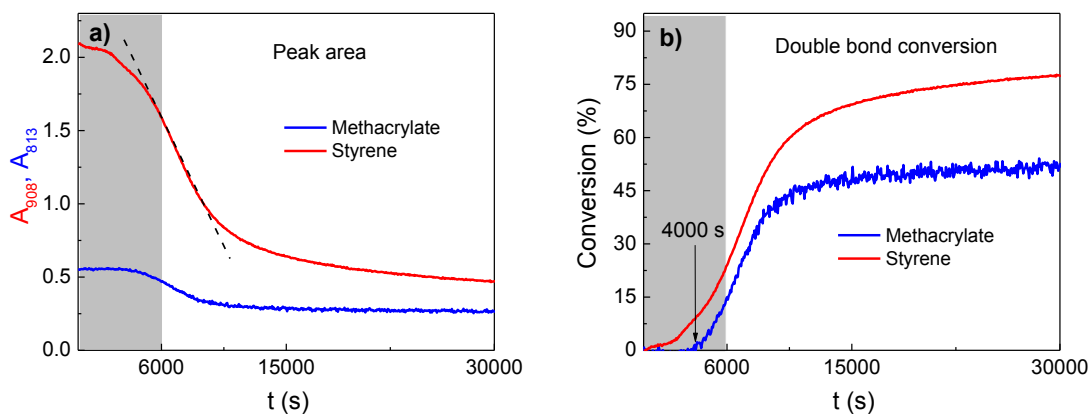

Figure 13: a) Evolution of the IR peak area at 908 and 813 cm$^{-1}$ reflecting the double bond consumption of respectively styrene (red line) and the methacrylate groups of the oligomer (blue line) of the VER (with intitiator) over a period of 10h at 23°C. The line represents the maximum slope of the area change at the inflection point of the curve. b) Same data expressed in a double bond conversion scale. The shaded area marks the time range of $|G^*|$ measurement of Fig. 12 i.e. 0 to 6000 s. The arrow indicates the point (4000 s) where methacrylate double bonds start to be consumed.

Figure 13(a) shows the evolution of the peak areas $A_{908}$ and $A_{813}$, characteristic for the double bonds of styrene and methacrylate respectively, over a period of 30000 s ($\approx$ 8.3 hours) for the VER crosslinking at 23°C. Most of the polymerization is clearly completed in less than half of this time, with a marked step between 3000 and 12000 s both for the styrene and methacrylate double bonds. The slope of the straight line at the inflection point of $A_{908}$ (= - 2 $10^{-4}$ s$^{-1}$) defines an upper bound for the gradient of the peak area which serves as analytical parameter for the chemical conversion. Interestingly, a spectroscopic mutation time can be introduced as $\tau_{Mu}^{S}(\bar{v}, t) = A_{\bar{v}}(t) / \frac{\partial |A_{\bar{v}}(t)|}{\partial t}$ in a generalization of Eq. (1) toward time-resolved spectroscopic experiments. For the present experiment, $\tau_{Mu}^{S}(908\ cm^{-1}, t_{inf}) \approx 6000$ s at the inflection point, $t_{inf}$. The corresponding mutation number then amounts to $N_{Mu}^{S} \approx 0.003$ which means that every averaged spectrum was recorded in quasi-steady conditions in this experiment. Figure 13(b) shows the same information with panel (a) but now expressed as relative conversion for the two different double bonds. Whereas styrene starts to polymerize immediately, a delay is observed for the consumption of the methacrylate groups with an onset at 4000 s, as pointed by the arrow in Figure 13(b). Given that methacrylate double bonds drives crosslinking and, hence, network buildup upon polymerization, it is not a surprise to find a correlation with the behavior of $|G^*|$ in Figure 12 where, there is a stronger growth after 4000 s consistent with the onset of methacrylate consumption shown in Figure 13(b). At earlier crosslinking times (<4000 s), the increase of $|G^*|$ essentially reflects the viscosity increase of the liquid resin owing to the prevalent polymerization of styrene as linear chains. The shaded area indicates the crosslinking time range of level of conversion for styrene ($\approx$ 19%) and methacrylate ($\approx$ 10%) at the end of the run shown in Figure 12. After 24h, the degree of conversion reached 84% and 53% for styrene end methacrylate respectively.

*Curing of Vinyl-ester resin with inorganic fillers*

The addition of inorganic fillers is a common practice to strengthen the mechanical properties of resin-based composites. A popular filler is hydrophilic fumed silica R200 (Evonik Industries AG); an industrial product with a nominal surface area of 200 m$^2$ g$^{-1}$ and high polydispersity of particle size. Primary particle diameter is 12nm (Evonik Bulletin) while aggregates range from 100 to 200 nm. These aggregates tend to form clusters that can break down and reform under flow, hence fumed silica dispersions exhibit thixotropic behavior while the LVE properties depend strongly on preparation history. The mechanical properties of the cured, as well the uncured sample, are greatly affected by the dispersion quality of the filler. The strain history and sample thixotropy is important in such filled sample and this cannot be fully controlled during loading and mixing affecting data reproducibility (Dullaert and Mewis 2005, Raghavan and Khan 1995). Fumed silica was dried in vacuum and elevated temperatures (80°C) overnight. The initiator was systematically introduced in the resin using a roller mixer for 180 s. Then fumed silica was added and dispersion was completed by means of a vortex mixer for another 180 s.

Reinforcement increases the modulus and response signal strength therefore measurements via an OWCh are possible even at the very early stages of crosslinking. A sequence of 16 OWCh-MCR702 chirps with $t_{exp}$ = 9 s was applied to the vinylester resin loaded with 8 w/w% fumed silica. At this concentration and once the mixing is completed the sample already exhibits a predominantly solid like response (gel). The LVE evolution during cure is shown in Figure 14 where the first measurement of the modulus at 2500 s is almost one order of magnitude higher than pristine vinylester resin of Figure 12. At early times, t < 4000 s, colloidal interactions are predominant. A space-filling network of flocs is formed due to interparticle attractions (Barthel 1995) and the modulus evolution is weakly affected by the slow increase of the matrix viscosity with the crosslinking rate significantly reduced due to reinforcement (Nikzamir et al. 2019). At longer times the matrix crosslinking leads to a dramatic increase of the elastic modulus. Given the rheological complexity of this mutating material with an evolving chemical network and ageing a systematic study is needed to explore the plethora of interactions and microstructural changes under flow. This was not the intended focus of the present work.

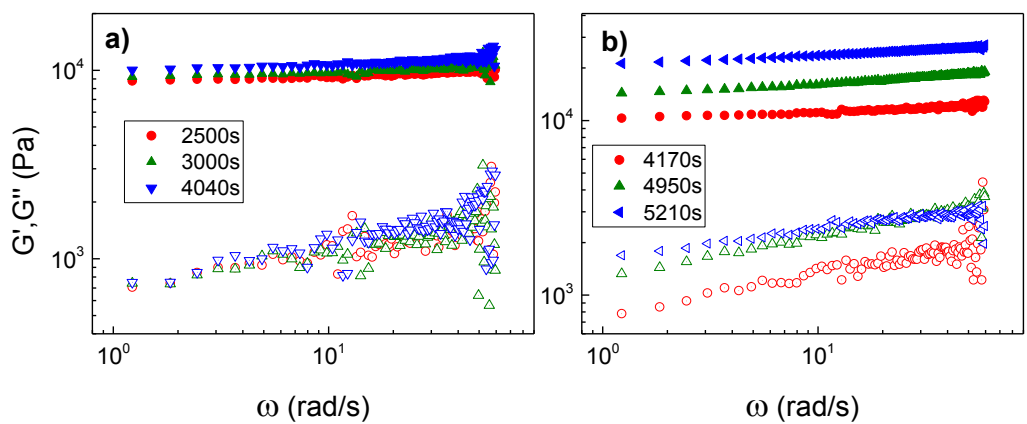

Figure 14. LVE spectra of curing VER and fumed silica R200 nanocomposite. Silica concentration is 8 w/w%. Measurements were performed with the OWCh-MCR702 technique with PP50mm geometry at various curing times and T=23 °C. Three representative spectra of the 16 measurements performed are shown for clarity at a) early and b) long times. Legend indicates the curing time (elapsed time from MEKP addition and mixing).

**CONCLUSIONS**

Undoubtedly, amplitude-frequency modulated chirp rheometry reduces the test time considerably compared to conventional DFS, hence it is the method of choice for the LVE interrogation of fast evolving systems. Application of the optimal windowed chirp (OWCh) in commercial rotational rheometers is straight forward though it requires protocols that are tailored to the specific instrument. Limitations are mainly imposed by the data acquisition system of the rheometer and the motor-controller's ability to impose the commanded strain waveform. These limits were exploited for the Anton Paar MCR702 and the simpler PZR and in each case the operability window and optimum range of parameters were discussed. Careful comparison with DFS data showed that the torque amplitude needs be increased at least 10 times more than in a conventional DFS for the MCR702. The PZR is less affected by data acquisition limitations and is capable of producing interrogating exponential chirp pulses as short as 1s, enabling us to keep $N_{Mu}$ sufficiently low, even for rapidly evolving systems. The technique was validated for both instruments and applied to follow the time-evolving rheology of a curing vinylester resin. The LVE evolution was measured in conjunction with styrene consumption. The mutation time varies considerably with the extent of cure hence a safe margin should be adopted. The two exponential regimes in the evolution of the complex modulus magnitude with crosslinking time, were explained by following styrene and methacrylate bond formation rate, by means of FTIR analysis. Confinement effects were found to be critical in gaps smaller than 0.3mm and such effects need to be considered when data are evaluated and modeled. Now that the proof of concept has been established, a systematic study with industrial resins reinforced with various fillers such as fumed silica will provide valuable information on the curing process of many commercial products. The toolbox developed in the present work can be utilized for studying a plethora of curing and time-evolving materials in the context of both fundamental and application-driven research.

**APPENDIX A**

The excitation strain waveforms (chirps) for both MCR702 and PZR were constructed based on the following equation:

$$\gamma(t) = \begin{cases} = \cos^2\left[\frac{\pi}{r}\left(\frac{t}{t_{exp}} - \frac{r}{2}\right)\right] f(t), & \text{for } \frac{t}{t_{exp}} \leq \frac{r}{2} \\ = f(t), & \text{for } \frac{r}{2} < \frac{t}{t_{exp}} < 1 - \frac{r}{2} \\ = \cos^2\left[\frac{\pi}{r}\left(\frac{t}{t_{exp}} - 1 + \frac{r}{2}\right)\right] f(t), & \text{for } \frac{t}{t_{exp}} \geq 1 - \frac{r}{2} \end{cases} \quad (A1)$$

where the function $f(t)$ corresponds to the non-tapered exponential up-chirp:

$$f(t) = \sin\left\{\frac{\omega_{min}}{\log\left(\frac{\omega_{max}}{\omega_{min}}\right)} t_{exp} \left[\exp\left(\log\left(\frac{\omega_{max}}{\omega_{min}}\right)\frac{t}{t_{exp}}\right) - 1\right]\right\}$$

(A2)

Equation 6 is the tapering function where the tapering ratio $r$ is set to 0.1 based on (Geri *et al.* 2018).

**APPENDIX B**

*Sensitivity to gap dimension*

When the gap decreases to values comparable to the characteristic length scale of the microstructure the continuum assumption breaks down and the measurement becomes sensitive

Page 32

to microstructural details, the so-called confinement effect. The PZR operates at gaps of 150 μm to mitigate sample inertia while a typical gap for plate-plate measurement in a rheometer ranges between 0.5 and 1.5 mm. Large gaps are limited by the distortion of the air-sample interface (when hydrostatic pressure overcomes surface tension) as well as sample inertia (Schrag 1977). On the other hand, small gaps amplify parallelism errors and enhance the confinement effect as the crosslinking sample may not be micro-structurally homogeneous (Clasen and McKinley 2004). As an example of this confinement effect we note prior studies of carbopol gels where data were affected at gaps smaller than ~150 μm (Liu *et al.* 2018, Yan *et al.* 2010) yet these studies concern nonlinear deformations. Carbopol is microgel dispersion rather than a chemical network nonetheless, the formation of interconnected clusters comparable to gap dimensions provides a direct path for stress transmission thus affecting the measurement. The presence of dispersed filler component in a rapidly polymerizing matrix phase may also give rise to heterogeneous stress transmission pathways at small length scales. We investigate this possibility here.

The effect of the characteristic gap dimension on the measured LVE of a crosslinking system stems from two major causes: the sensitivity of the followed crosslinking path to sample dimension (thickness) and the confinement effect. To verify this we performed different runs of Dynamic Time Sweeps (DTS) in the MCR702 with parallel plates of 25mm radius. The gap was varied from 1.2 to 0.15 mm. To rule out the possibility of an alignment error or the finite optical encoder angular deflection resolution becoming important in small gaps, we performed a test run with a pDMS viscoelastic standard with the same gap values. No gap dependence was observed as shown in Figure S6 of the supplemental information.

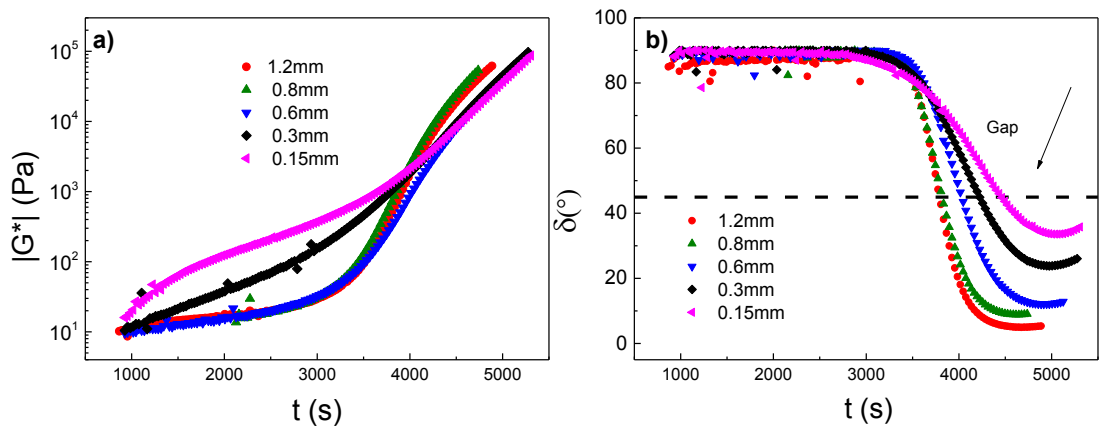

Figure 15. Gap dependence of a) $|G^*|$ and b) phase angle $\delta$ on crosslinking time for different sample thicknesses as indicated in the legend. Dynamic time sweeps were performed with MCR702 and PP25 geometry at 53.2 rad/s and T=23 °C. Dashed line indicates the liquid to solid-like transition ($\delta = 45°$).

However, the results from the crosslinking VER presented in Figure 15 confirm a strong dependence of both the polar components of $G^*$ on the rheometric gap. At early times (up to 3500 s) the reduction of gap from 1.2 to 0.6 mm has no effect on the evolution of $|G^*(t;\omega)|$. Minor differences can be attributed to small (composition-dependent) variations in crosslinking path. When the gap was further reduced to 0.3 and 0.15mm, $|G^*|$ exhibited faster increase at early times, presumably due to the confinement effect. Once the network was fully established (t > 4000s), $|G^*|$ evolution was again found to be similar for all gaps. The phase angle dependence is initially weak at early times however there is a strong increase at smaller gaps in the post gel regime. This can arise due to both confinement effects and slower crosslinking as supported by the lower values of $|G^*|$ for smaller gaps. This gap dependence could in principle be hypothesize to result from wall slip effects (Yoshimura and Prud'homme 1988) however this can be ruled out in this material during its curing stage as the sample is literally glued to the tools.

*Compliance effects*

Instrument torsional compliance could also affect the fidelity of the measurements as the gap is reduced and this contribution cannot be totally excluded. However, we have additional technical reasons to support the conclusion that the observed gap dependency cannot be attributed to torsional compliance. These are: a) the instrument's torsional compliance should be predominantly elastic (i.e. in phase with deformation). When data are affected by this contribution the phase angle is typically underestimated (Sternstein 1983). However, in Figure 15(b) the phase angle increases as the gap is decreased (smaller gaps should have larger compliance effect). b) The effects of torsional compliance on $|G^*|$ should be stronger at late crosslinking times, when the material itself is stiffer. On the contrary Figure 15(a) shows that the effect is stronger at early times where $|G^*| < 10^3$ Pa. Moreover at approximately same curing time $|G^*|$ is almost unaffected for gaps 1.2, 0.8 and 0.6 mm. The empirical evidence shows it becomes sensitive to gap dimensions for gaps < 0.6mm. When torsional compliance dominates $|G^*|$ is underestimated (Liu *et al.* 2011) while in Figure 15 it is overestimated. This is a signature of "stronger" stress transmission pathway at small gaps. To further support this argument we performed a similar experiment with a homogeneous reference sample (pDMS) at similar gap range. From Figure S6 (see SI) it is evident that the LVE spectra of this reference sample exhibits no gap dependence even for $|G^*| > 10^3$ Pa.

Another compliance effect stems from instrument's axial compliance when normal forces are present. Axial compliance can alter the gap from its predetermined value and induce errors in the measured shear stress amplitude (Schweizer *et al.* 2004). This is clearly not dominant in the data of Figure 15(a) as the strongest gap dependence of $|G^*|$ is observed at the early crosslinking times when normal forces are weak.

To summarize, the size of the gap clearly affects measurements of the time evolving viscoelastic moduli in this rapidly crosslinking material and this should be accounted for when comparing data from different instruments and geometries. Small gaps promote increasingly stiff, solid-like response as stresses are transmitted across the sample via shorter paths. At a constant gap the temporal evolution of the phase angle $\delta(t)$ shown in Figure 15(b) exhibits a non-monotonic behavior. For all gaps at early times, $\delta$ decreases initially towards the liquid to solid transition ($\delta = 45°$) however, it subsequently exhibits an upturn at long times suggestive of increased dissipation mechanisms in the crosslinking system, confirming the trend observed in OWCh-PZR measurements (see Figure 11). This inflection point has been attributed to the onset of vitrification processes (Aoki *et al.* 2019, Harran and Laudouard 1986, Lange *et al.* 1999) and the presence of dangling chains. Chemical cross-linking also promotes denser packing and thus volume shrinkage which leads to negative normal forces exerted by the sample if is held at constant gap during cure. To circumvent difficulties arising from instrument compliance a zero normal force protocol is utilized by many authors (Lehéricey *et al.* 2021). In the PZR such a protocol is not feasible. In an effort to ensure similar experimental conditions in both instruments a zero normal force condition was not imposed in the MCR702. Measurements were performed until the axial forces reached 5N and then terminated (Fig S5). Small gaps and small sample volume mitigate the magnitude of the normal forces exerted as the total normal displacement is reduced during the crosslinking.

**APPENDIX C**

*List of symbols*

| Symbols | Comments |
|---|---|
| $f_s$ | Sampling frequency |
| $M$ | Torque in rotational rheometers |
| $M_o$ | Torque amplitude |
| $N$ | Number of samples in time domain |
| $N_f$ | Number of samples in frequency domain |
| $N_{Mu}$ | Mutation number |
| $t_{exp}$ | Total duration of the measurement |
| $t_n$ | Time step |
| $t_{osc}$ | Period of oscillation |
| $t_{sweep}$ | Time to complete a Dynamic Frequency Sweep |
| $\Delta\omega$ | Frequency step |
| $\tau_{Mu}^{|G^*|}$ | Mutation time with respect to $|G^*|$ |
| $\tau_{Mu}^{S}$ | Spectroscopic Mutation time |
| $\omega_{min}$ | The lowest frequency of chirp |
| $\omega_{max}$ | The highest frequency of chirp |

**ACKNOWLEDGEMENTS:** We thank N. Kalafatakis for his help with measurements during the early stages of this study. We acknowledge B. Keshavarz for his assistance in the implementation of the original chirp protocol.

# AUTHOR DECLARATIONS

## Conflict of interest

The authors have no conflicts to disclose.